\documentclass[preprint]{aastex}

\newcommand{\unit}[1]{\; \mbox{#1}}
\newcommand{\un}[1]{_{\mbox{\scriptsize #1}}}

\usepackage{color}
\usepackage{epsfig}
\usepackage{amssymb}

\begin{document}



\title{Cosmic Rays below $Z=30$ in a diffusion model: new
constraints on propagation parameters.}
\author{D. Maurin}
\affil{Laboratoire de Physique Th\'eorique {\sc lapth}, 
Annecy--le--Vieux, 74941, France \\
Universit\'e de Savoie, Chamb\'ery, 73011, France}
\email{maurin@lapp.in2p3.fr}
\author{F. Donato\altaffilmark{1}} 
\affil{Laboratoire de Physique Th\'eorique {\sc lapth}, 
Annecy--le--Vieux, 74941, France}
\email{donato@lapp.in2p3.fr}
\and
\author{R. Taillet and P.Salati}
\affil{Laboratoire de Physique Th\'eorique {\sc lapth}, 
Annecy--le--Vieux, 74941, France \\
Universit\'e de Savoie, Chamb\'ery, 73011, France.}
\email{taillet@lapp.in2p3.fr, salati@lapp.in2p3.fr}
\altaffiltext{1}{{\sc infn} post--doctoral Fellow}




\begin{abstract}

Cosmic ray nuclei fluxes are expected to be measured with high
precision in the near future. For instance, high quality data on the
antiproton
component could give important clues about the nature of the astronomical
dark matter.
A very good understanding of the different aspects of cosmic
ray propagation is therefore necessary.
In this paper, we use cosmic
ray nuclei data to give constraints on the diffusion parameters.
Propagation is studied with semi--analytical solutions of a diffusion
model, and we give new analytical solutions for radioactively produced
species.
Our model includes convection and reacceleration as well as the
standard energy losses.
We perform a $\chi^2$ analysis over B/C data  for a large number of
configurations obtained by varying the relevant parameters of the diffusion
model.
A very good agreement with B/C data arises for a number of
configurations, all of which are compatible with sub--Fe/Fe data.
Different source spectra $Q(E)$ and diffusion coefficients $K(E)$ 
have been tried,
but for both parameters only one form gives a good fit.
Another important result is that models without convection or without
reacceleration are excluded.
We find that the various parameters, {\em i.e.} the diffusion
coefficient normalisation $K_0$ and spectral index $\delta$, 
the halo thickness $L$, 
the Alfv\'en velocity $V_a$, and the convection velocity $V_c$
are strongly correlated.
We obtain limits on the spectral index $\delta$ of the
diffusion coefficient, and in particular we exclude a Kolmogorov
spectrum ($\delta = 1/3$).
\end{abstract}



\section{Introduction}
Understanding the composition and spectral features of cosmic rays has
always been an astrophysical challenge.
On one hand, the observational data have long been scarce and suffered
from large uncertainties. On the other hand, the theoretical predictions
to which these data should be compared to have also suffered from
several drawbacks.
Composition and spectra arise from the nuclear 
interaction of an initial
distribution of energetic particles with interstellar matter
({\em spallations}) and their electromagnetic interactions with
galactic magnetic fields ({\em acceleration} and {\em diffusive
reacceleration}).
First, the nuclear cross sections to be used were not very well known
until recently.
Second, our knowledge of the galactic magnetic field is far from
complete. Cosmic rays are sensitive to its scale inhomogeneities
({\em diffusion}) which are not well observed.
They are also sensitive to the presence of plasma shock--waves
(acceleration in localized sources and diffusive reacceleration).
Third, composition and spectra are altered as the cosmic rays enter
the solar magnetic field, so that some more modelling has to be
done in order to infer interstellar spectra from observations.

However, despite these sources of uncertainty, some gross features of
the cosmic ray properties are well established.
First, the study of secondary--to--primary ratio shows that a cosmic ray
has crossed an average of about $x=9 \unit{g} \unit{cm}^{-2}$ of interstellar
matter between its acceleration and its detection.
Second, the isotopic ratio of radioactive species shows that it took
about $\tau = 20 \unit{Myr}$ between the same events. As cosmic rays have a
velocity close to $c$, we can infer the average density of the medium
to be $n= x/(\tau \,c\,m_{H}) \approx 0.3 \unit{cm}^{-3}$.
As the disk density is about $1 \unit{cm}^{-3}$, the cosmic rays must
spend a fraction of time in an empty region, called the {\em diffusion
halo}. Cosmic rays are produced and destroyed in the
galactic disk, they diffuse in a larger zone whose characteristics are
not well known, and eventually they can escape from this zone.
To have a good modelling of cosmic ray propagation, we should know
the geometry and size of the diffusion zone, the characteristics of the
galactic magnetic field, and the sources.

In this paper, we use the existing data on
nuclei, gathered by balloon--borne and space experiments during the last
thirty years, to put constraints on the parameters describing the
propagation of cosmic ray nuclei.
One consequence is that many of the large uncertainties affecting the
calculation of the antiproton flux could be strongly reduced (F. Donato \& al.,
in preparation  \cite{la donato}).
This is of utmost relevance for the study of exotic sources of
antiprotons and antideuterons.


\section{The diffusion model}

It has been recognized for a long time that the relevant physical
propagation model to be used is the diffusion model (Berezinskii \& al. 1990;
Ginzburg \& Syrovatskii 1964), though the so--called leaky box model has been
widely preferred for decades because of its simplicity.

The steady--state differential density $N^j(E,\vec{r})$ of the nucleus $j$ as a
function of energy $E$ and position $\vec{r}$ in the Galaxy,
is given by (see for instance Berezinskii \& al. 1990)
\begin{eqnarray}
         \label{EQUATION LA PLUS GENERALE}
         \nabla . (K^j\nabla N^j - V_c N^j)
         -\frac{\partial}{\partial E}\left( \frac{\nabla.V_c}{3}
	 E_k\left(\frac{2m+E_k}{m+E_k}\right)N^j\right)
         +\frac{\partial}{\partial
E}(b^jN^j)-\frac{1}{2}\frac{\partial^2}{\partial
         E^2}(d^jN^j)+\tilde{\Gamma}^j N^j=
         q^j+\!\!\sum_{m_k>m_j}\!\tilde{\Gamma}^{kj}N^k
\end{eqnarray}
The first terms represent
diffusion ($K^j$ is the diffusion coefficient)
and convection ($V_c$ is the convection velocity). 
The divergence of this velocity, expressed in the next term
gives rise to an energy loss term connected with
the adiabatic expansion of cosmic rays.
Further, we have to take into account
ionization and coulombian losses (the energy loss coefficients
are specified below), plus
a reacceleration term in first order derivative (all included in
$b^j$) and finally a second order derivative in $E$ for the associated
second order term in reacceleration ($d^j$ is the energy diffusion
coefficient).
These stand for the continuous losses.
The last term of the l.h.s. takes care of the disappearance of the nucleus
$j$ ($\tilde{\Gamma}^j$ for short)
due to its collisions with interstellar
matter ({\sc ism}).
In the r.h.s., the source term $q^j$ takes
into account
the primary production and acceleration of nuclei described by
an injection spectrum
(for the sake of clarity, we have not written down the terms
describing the contribution of radioactive species).
Finally, the last term is for the secondary $j$ sources,
namely spallation contribution\footnote{We use here the
so--called straight--ahead approximation which implies the conservation of
kinetic energy per nucleon during a spallation process
\begin{equation}
\int_0^{\infty}n_H v' N^k(T')\sigma^{kj}(T,T')dT'=
\int_0^{\infty}n_H v' N^k(T')\sigma^{kj}(T)\delta (T-T')dT'=
n_H v N^k(T)\sigma^{kj}(T)
\end{equation}
}
$\tilde{\Gamma}^{kj}$ from all other
heavier nuclei. We use here the notation $E$ for total energy
and $E_k$ for kinetic energy.

One needs to solve a complete  triangular--like set of coupled equations since
only heaviest nuclei contribute to a given nucleus.
Quantities in this equation are functions of spatial coordinates
(not time, steady--state being assumed) and of energy.
Unless stated otherwise, the word {\em energy} will stand for {\em
kinetic energy per nucleon} ($T$),
since this is the appropriate parameter to be used, as it is conserved in
spallation reactions (see footnote~2).

The other very popular model is the leaky box, in which all the quantities are
spatially averaged (so that convection has no meaning),
and the diffusion term is then replaced by an escape term
\[
      -\nabla . (K^j\nabla N^j)\longleftrightarrow
       \bar{N^j}/\tau_{esc}
      \]
      which has the
      meaning of a residence time $\tau_{esc}$ (Myr) in 
      the confinement volume.
Note that in this case, the mean time $\langle \tau \rangle$
spent in the box and
the mean column density $\langle x \rangle$
(in units of g$\unit{cm}^{-2}$) of crossed matter 
are treated on the same footing since the model is homogeneous.
This is not true for multi--zone models and especially for the
diffusion model where cosmic rays
spend most of their time in the diffusion halo which has a zero matter density.
It may seem very surprising that this simplified model is able to
reproduce many observations on stable nuclei.
Actually, it can be shown that leaky box models are  
often equivalent to diffusion models, as far
as stable species are concerned (for a generic 
discussion on the ``leakage--lifetime" approximation in cosmic ray diffusion,
see Jones 70).
This is not obvious from the equations, but it can be seen
more readily if we solve them in a formalism called the weighted slab
technique.
Basically, it consists in writing a general solution under the form
(see for example Berezinskii et al. 1990; Ginzburg \& Syrovatskii 1964)
\begin{equation}
         N^j({\bf r})=\int_0^{\infty}{\tilde N}^j (x)G({\bf r},x)dx
         \label{def: WSM}
\end{equation}
Thus, (\ref{EQUATION LA PLUS GENERALE})
decouples into two independent equations. The first one involves 
$G({\bf r},x)$. It depends only on the geometry of the problem and on 
the chosen diffusion scheme (or the leaky box parameters),
but not on the particular species $j$. 
The other one involves ${\tilde N}^j (x)$ and contains the nuclear physics 
aspects of the propagation.
Different models (diffusion or leaky box) correspond to different 
$G({\bf r},x)$,
but the equations on ${\tilde N}^j (x)$ are the same.
$G$ is the path length distribution
and is interpreted as the distribution of probability
that a nucleus $j$ goes through a column density $x$ before
reaching Earth.
The global solution is then obtained by a convolution of the
surviving quantity ${\tilde N}^j (x)$ under nuclear processes
after $x \unit{g}\unit{cm}^{-2}$, with the
weighted probability associated.

In a leaky box model, it can be shown that the normalized path length
distribution function
is given by
      \begin{equation}
      G(x)=\frac{1}{\lambda\un{esc}}\exp \left( 
      \frac{-x}{\lambda\un{esc}}\right)
      \end{equation}
      where the average quantity of matter crossed by a cosmic ray is given by
      \begin{equation}
        \left< x \right>= \int_0^{\infty} xG(x)\;dx =\lambda\un{esc}
      \end{equation}
In the diffusion model, it has been shown (Owens 1976) that for a wide class
of geometries,
the function $G$ is given by an infinite series of exponentials
involving the diffusion coefficient.
Moreover, the physical parameters are such that for sufficiently
large grammage $x$,  only the first exponential contributes 
so that we recover a leaky box model (Jones 1970; Ginzburg, Khazan \& Ptuskin 1980; Schlickeiser \& Lerche 1985; 
Lerche \&  Schlickeiser 1985). For the model used in this paper, 
it can be shown that $\lambda\un{esc}=\bar{m}nvhL/K$
in the limit $L\ll R$.
However, this is no longer true for lower $x$, and then the leaky box
model fails.
This may explain, in conjunction with poorly known cross sections
for heavy nuclei, why for many years cosmic ray physicists had to use
modifications (Lezniak \& Webber 1979; Garcia--Munoz \& al. 1987; 
Webber \& al. 1998b) of the
leaky box path length distribution at small grammage to reconcile 
observations of B/C with
those of sub--Fe/Fe.
Even worst, this approximate equivalence between leaky box and
diffusion model breaks down for radioactive species 
(Prishchep \& Ptuskin 1975).

This motivates  us to choose the diffusion model and solve the corresponding
equation~(\ref{EQUATION LA PLUS GENERALE}).
From a more general point of view, note also that
the weighted slab technique mentionned above should be used carefully.
Actually, it
was recently realized (Ptuskin, Jones \& Ormes 1996) that pre--1996
estimations of cosmic ray parameters with this technique
were not exact.
First, at fixed energy per nucleon, the rigidity and therefore the propagation
features depend on the nuclear species, whatever the energy. 
This prevents the above--mentionned decoupling.
Second, energy losses that strongly depend on $A$ and $Z$ also
take place at low energy. 
Ptuskin et al. 1996 presented a modification of
this formalism which is correct in both the highly and non--relativistic 
regimes. These fundamental findings were confirmed 
by a more complete numerical treatment  (Stephens \& Streitmatter 1998) 
and give rise to differences of as much as 20\%.

For the nuclear part, we
adopted the direct approach of the shower technique.
This means that the flux is first evaluated for the heavier primary cosmic ray,
for which the diffusion equation is simple and does not couple to any
other species.
Then, the flux of the next nucleus is computed, with a spallation term
depending only on the heavier nucleus, whose flux is known from the
previous step.
This procedure is repeated for all the nuclei down to the lightest
one.
We started at $Z=30$, the heavier non negligeable primary
found in cosmic radiation (Binns \& al. 1989),
since $(Z>30)\leq 10^{-3}$ Fe.
For all light nuclei, we checked that it is sufficient to start 
from $Z=16$ (S).
For the detail of solutions, the reader is referred to
Appendix~\ref{sec: sol diffusion model}.
The next section is devoted to the inputs of the model.


\section{The parameters of the model}
        \subsection{Geometry of the Galaxy}
The geometry of the problem used here is a classical cylindrical
box (see for example Webber et al. 1992) whose radial extension is $R$,
with a disk of thickness $2h$ and a halo of half--height $L$.
Only $L$ is not fixed
(see $\S$\ref{sec: results}) and $R=20\unit{kpc}$, $h=100\unit{pc}$.
Sources and interactions with matter are confined to the thin disk and
diffusion which occurs throughout disc and halo with the same strength
is independent of space coordinates.
The solar system is located in the galactic disc ($z=0$) and at a
centrogalactic distance $R_\odot=8 \unit{kpc}$ 
(Stanek \& Garnavich 1998; Alves 2000).
        \subsection{Cross sections}
        Cross sections play a crucial role in propagating the cosmic rays
        throughout the Galaxy. We have to distinguish bewtween
        the total inelastic cross section and
        the spallation cross section.
          \subsubsection{Total inelastic cross section}
            \label{sec: reaction cross section}
          The total inelastic cross section, which is actually
          a reaction cross section, is defined by
            \begin{equation}
            \sigma_{inel}^{tot}=\sigma^{tot}-\sigma^{tot}_{elastic}
            \end{equation}
          Various empirical modifications of the original (Bradt \& Peters
          1950) geometrical approach
          have been used to parameterize this total cross section.
          Letaw, Silberberg \& Tsao (1983) and  Silberberg \& Tsao (1990) produced
          the basic energy dependent equations further refined in Sihver
          \& al. (1993) and in Wellish \& Axen (1996).
          Recently Tripathi, Cucinotta \& Wilson (1997a, 1997b, 1999)
	  proposed a universal parameterization which is valid for all
          nucleus--nucleus reactions, including neutron induced reaction.
          All of these approaches have been compared in
          Silberberg, Tsao \& Barghouty (1998), where the preference
          were given to Tripathi et al. (1997a, 1997b, 1999).
          For this reason, we adopted the latter parameterization throughout
          this study. The precision can be estimated to be about $5\%$.
          \subsubsection{Spallation cross section}
             \label{sec: spallation reaction}
            The fragmentation cross sections
            are somewhat more complicated. Basically, we can separate
            three distinct approaches~: (i) microscopical
            description (Bondorf \& al. 1995; Ramsey \& al. 1998)
	    (ii) semi--empirical formulae
	    (Silberberg et al. 1998; Tsao, Silberberg \& Barghouty, 1998;
	    Tsao et al. 1999) and (iii) empirical
            formulae (Webber, Kish \& Schrier 1990d). The first one
            takes into account the fundamental nuclear physics
            but is very computer time consuming
            and has not reached a precision as good as the two other
             approaches. The second method takes into
            account more phenomenological inputs, and is
            better designed for the unmeasured fragmentation cross sections
            which correspond mainly to the ultra heavy ($Z\geq 30$) nuclei.
            The third one breaks down in this region, but can
	   be used for $3\leq Z\leq 30$ because it is adapted
	   to fit measurements. Consequently, we used the code of
           Webber et al. (1990d) available on the 
	   web~\footnote{http://spdsch.phys.lsu.edu/SPDSCH\_Pages/Software\_Pages/Cross\_Section\_Calcs/CrossSectionCalcs.html}
            updated with new parameters (table~V of Webber \& al. 1998a).
            This parameterization takes
            advantage of the fact that about $98\%$ of all the reaction involved
            in cosmic ray propagation have now at least one point of energy
            measured. This set of formulae gives a precision of about $10\%$
            but can not be further refined (Webber et al. 1998a).
            The code is extended for spallation on He with the parameterization
            given in Ferrando et al. (1988)\footnote{The parameterization 
	    given in this paper should be modified at low energy
	    and for large $\Delta Z$ (P. Ferrando 2001, 
	    private communication).}.

            To go further than this
            treatment, one should keep in mind the following points.
          First, the process of single or double nucleon removal is an
          important channel for the production of numerous nuclei.
          It has been parameterized in detail in 
	  Norbury \& Townsend (1993) and in Norbury \& Mueller (1994),
	   although in the following we will consider
	  Webber's approach sufficient for our purpose.
            Second, most of the parameterizations make use of the 
	straight--ahead approximation (see footnote~2).
         Tsao \& al. (1995) relaxed this
            condition and find a 5\% effect around the $1\unit{GeV}$ peak of
	   B/C
            (note that this effect is negligable for primary to primary and
            secondary to secondary ratios).
            Finally, it was found that all the measurements are not always
            consistent and systematic effects are likely to be present
         (Taddeucci \& al. 1997; Vonach \& al. 1997; Zeitlin \& al. 1997;
	 Flesh \& al. 1999; Korejwo \& al. 1999). 

           To summarize, we consider that Webber's code
            is very well suited for modelling spallation cross sections,
            thanks to the many experiments developed these last
	   years (Webber, Kish \& Schrier 1990a, 1990b, 1990c; 
	   Webber et al. 1998b, 1998c; Chen et al. 1997).
            The overall precision  can be roughly estimated to be better
		than 10\% on H, and better than 20\% on He.
        \subsection{Nuclear parameters}

        Apart from the stable species, we can distinguish
	two  categories of nuclei to be propagated, 
	depending on whether they live long enough to be
        observed or they decay quickly into other nuclei.
          \subsubsection{Ghost nuclei}

          Ghost nuclei are defined as those having a half--lifetime
	  larger than $\sim 1\unit{ms}$
          (so that they are detected in cross section measurements)
          and less than few hundred years (so that we can consider
          that disintegration is fast on a propagation
          timescale).
	 Several reaction channels may yield a given nucleus.
          This means that for this nucleus, we have to add to the
          direct production
          the contributions of all the short lived parent nuclei
            \begin{equation}
            \sigma^{effective}_{k\rightarrow j}=\sigma_{kj}+
            \sum^{\mbox{\scriptsize{\em ghost}}}_{nuclei \;\; X}
	    \sigma_{k\rightarrow X}{\cal B}r(X \rightarrow j)
            \end{equation}
            where ${\cal B}r(X\rightarrow j)$ is the branching ratio of
            the corresponding channel\footnote{To give an example, 
such a reaction
          reads for $^{9}_4Be$
            \begin{equation}
            \sigma^{effective}_{k\rightarrow ^{9}_4Be}=
             \sigma_{i\rightarrow ^{9}_4Be}
                    + 49.2\%\;\;\sigma_{k\rightarrow ^{9}_3Li}
                    +4.1\%\;\;\sigma_{k\rightarrow ^{11}_3Li}
            \end{equation}
	    }.
	            A table of all the disintegration chains
          can be found in Letaw \& al. 1984.
          To see the effect of the recent discovery of
          new intermediate short--lived nuclei and of the changes in branching
          ratio, we reconstructed from a recent compilation of nuclear
          properties (Audi et al. 1997) all the ghost nuclei and compared 
	  them with the Letaw table. We found that the net effect 
	  is negligible.
          This is because the discovered radioelements are very far from the
          valley of stability and their production is
          suppressed (as it can be observed experimentally). Such 
	  reaction chains should accordingly be taken as a fixed input 
	  for the future.

          \subsubsection{Radioactive nuclei}
          Radioactive nuclei can be separated into three groups~: unstable under
          $\beta$ decay, unstable under electronic capture and unstable under
          both reactions. All the physical effects related to these
          nuclei will be analysed in detail in a separate paper (F. Donato \&
	  al., in preparation \cite{la donato2}).

\subsection{Diffusion coefficient}
\label{sec: diffusion}
Physically, diffusion arises because charged particles interact with
the galactic magnetic field inhomogeneities.
This is an energy--dependent process because
higher energy particles are sensitive to
larger spatial scales. As a result, the diffusion
coefficient is related to the power spectrum of these inhomogeneities,
which is poorly known.
Several analytical forms of this energy dependence have been assumed
in the literature.
In particular, leaky box models give a purely phenomenological form
\begin{displaymath}
      \label{Forme lambda}
       \left\{
        \begin{array}{ll}
          \lambda\un{esc}=\lambda_0\beta &\mbox{if ${\cal R}<{\cal R}_c\unit{in GV}$}\\
          \lambda\un{esc}=\lambda_0\beta \left( {\cal R}/{\cal R}_c \right)^{-\delta} &\mbox{if
${\cal R}>{\cal R}_c\unit{in GV}$}\\
        \end{array}
        \right.
\end{displaymath}
where ${\cal R}=p/Z$ stands for the particle rigidity. 
A more fundamental origin may be found for the previous expression,
as one can show in a one--dimensional diffusion model with convection
(Jones \& al. 2001).
All the effects that led to that
phenomenological form are explicitely taken into account in 
our treatment and we may therefore
consider a more natural form for $K(E)$, inferred from
magnetohydrodynamics considerations (Ptuskin et al. 1997):
\begin{equation}
          K(E) = K_0 \, \beta \times {\cal R}^\delta
          \label{eq:diffusion_coefficient}
\end{equation}
where the normalisation $K_0$ is expressed in kpc$^2\unit{Myr}^{-1}$.
Nevertheless, we have also tested leaky box inspired forms, but these
never give good fits to the data (see next section).

\subsection{Sources}
\label{sec: source term}

A charged particle is called a cosmic ray when it has been
accelerated to an energy greater than $100 \unit{MeV/n}$ as far as our
study is concerned.
Several physical processes may be responsible for this energy gain; 
they all involve an interaction of the particle with a plasma shockwave
front.
This is responsible for the power--law energy dependence (see below).
The term {\em source} refers to the place where the charged particles
have been promoted to cosmic rays.
We present here the various hypotheses about their spatial
and spectral distributions and their compositions.

\subsubsection{Spatial distribution}

First, as explained above, we assumed that the Galaxy has a
cylindrical symmetry and that the sources are located in
the galactic disc.
So we only have to specify a radial distribution.
We have used the Case \& Bhattacharya (1996, 1998) 
distributions, but the results presented below 
are not sensitive to this particular choice.

\subsubsection{Spectral distribution}
\label{grumph}
The source spectral distributions can be split in two terms
\begin{equation}
          q^{j}(E) = q_0^j Q^j(E)
          \label{eq:source}
\end{equation}
where $q_0^{j}$ represent the composition and $Q^j(E)$ the energy
dependence.
Several spectra have been used in the past (Engelmann \& al. 1985),
\begin{eqnarray}
        Q^j(E) &\propto& p^{-\alpha_j} \nonumber\\
        Q^j(E) &\propto& \beta\, E_{kin}^{-\alpha_j}\\
        Q^j(E) &\propto& E_{tot}^{-\alpha_j}\nonumber
\end{eqnarray}
where $p$ is the momentum and where the spectral index $\alpha_j$ may depend
on the species $j$ considered. These forms are equivalent at high energies
but we focused on the first one only (equivalent to a power--law in
rigidity) as the others do not yield good fits to the data.
The sum $\delta+\alpha_j$ gives the overall power--law index of each
species at high energy ($\approx 1\unit{TeV}/n$): this is because 
the measured flux at these energy is
proportional to $Q^j(E)/K(E)\propto {\cal R}^{\delta+\alpha_j}$. In other 
words, at sufficiently high energy, energetic redistribution and 
spallation reactions do not affect propagation.
Consequently, measured spectral indexes enter as a fixed input, and
we took them in
Wiebel--Sooth, Biermann \& Meyer (1998).
For consistency, we required that whatever $\delta$,
the parameters $\alpha_j$ are adjusted so as to reproduce the
above--mentionned values.
This ensures that the spectra we compute
agree with data at high energy.

\subsubsection{Composition}
\label{sec:composition}

The composition $q_0^j$ is usually inferred from a comparison between
leaky box calculations and data.
Alternatively, a more fundamental approach can be used.
It may be taken either as the solar system composition convoluted with the
{\sc fip} (First Ionisation Potential), or convoluted with the
volatility.
As pointed out in Webber (1997), the two alternatives are not so different.
In this work, the isotopic composition of each element
was taken from Anders \& Grevesse (1989) and Grevesse \& Sauval (1998), 
and the $q_0$ dependence of
the {\sc fip} has been taken from Binns et al. (1989).
Then the primary cosmic ray
elemental composition has been adjusted to fit the {\sc heao}--3 data at
$10.6 \unit{GeV/n}$, while keeping constant the relative isotopic
abundances.

\subsection{Galactic wind, energy loss, reacceleration}
Once we have chosen a generic model,  we have to include
effects which affect the spectra at low energies.
A first step is the inclusion of energetic losses.
Then a most complete treatment must take into account reacceleration and
convection.
          \subsubsection{Energy losses}
            \label{sec: energy losses}
There are two types of energy losses which are relevant for nuclei:
ionization losses in the {\sc ism} neutral matter ($90$\% H and $10$\% He),
and Coulomb energy losses in a completely ionized plasma,
dominated by scattering off the thermal electrons.
We will use $<n_e>\sim 0.033\unit{cm}^{-3}$,
and $T_e\sim10^4\unit{K}$ (Nordgren 1992).
Complete formulae for these two effects
are compiled, for example, in Strong \& Moskalenko (1998) or in 
Mannheim \& Schlickeiser (1994). 
There is also a third type of loss 
related to the existence of a galactic wind.
We recall that in equation (\ref{EQUATION LA PLUS GENERALE}), there is a term
proportional to $\nabla.V_c$.
As we choose a constant wind velocity in the $z$ direction (see farther),
we could conclude at a first glance that this term vanishes.
Actually, this is the case everywhere in the Galaxy except at $z=0$ where a
discontinuity occurs, due to the opposite sign of the wind velocity 
above and below the galactic plane. One gets a term that can be
expressed in the same form as ionization and coulombian losses
with an effective term
\begin{equation}
\left< \frac{dE}{dt}\right>_{Adiab}=-E_k\left(\frac{2m+E_k}{m+E_k}\right)
\frac{V_c}{3h}
\end{equation}
$E_k$ stands for the total kinetic energy and it should not be confused
with the kinetic energy per nucleus frequently used in this paper.
Finally, because of these effects the propagation equation 
must be solved numerically.
          \subsubsection{Reacceleration}
            \label{sec: reacceleration}
As the fractional energy changes in a single collision are small, we can treat
Fermi acceleration (Fermi 1949; 1954) using the 
Fokker--Planck formalism (Blandford \& Eichler 1987).
In one dimension
\begin{eqnarray}
      \frac{\partial f}{\partial t}+v\frac{\partial f}{\partial x}&=&
	\frac{\partial}{\partial p}\left[
	     - \left< \frac{\Delta p}{\Delta t} \right> f +
	      \frac{1}{2}\frac{\partial}{\partial p}\left(
	      \left< \frac{(\Delta p)^2}{\Delta t}\right>f\right)\right]
   \end{eqnarray}
   where $\langle \Delta p/\Delta t \rangle$ and
   $\langle (\Delta p)^2/\Delta t\rangle$
are expressed in terms of the probability for changing the momentum
$p$ to $p+\Delta p$ during time $\Delta t$.
A simplification occurs when the recoil of the scatterer can be
ignored. In general, the principle of detailed balance ensures 
that the probability
for a gain is equal to that for a loss. This implies
    \begin{equation}
	\left<\frac{\Delta p}{\Delta t}\right> =\frac{1}{2} 
	\frac{\partial}{\partial p}
	\left< \frac{(\Delta p)^2}{\Delta t}\right>
    \end{equation}
which reduces the equation into
\begin{eqnarray}
    \label{detailed balance approx}
      \frac{\partial f}{\partial t}+v\frac{\partial f}{\partial x}=
	\frac{\partial}{\partial p}K_{pp}
	    \frac{\partial}{\partial p}\left(
	      f\right)
   \end{eqnarray}
where
\begin{equation}
K_{pp}\equiv \frac{1}{2}
	\left< \frac{(\Delta p)^2}{\Delta t}\right>
\end{equation}
 From another point of view, we can start
with a collisionless Boltzmann equation
and expand it up to second order in perturbed quantities
(magnetic irregularities). It gives the full diffusion equation
in quasi--linear theory (Schlickeiser 1986; Kulsrud \& Pearce 1969). Assuming a
constant flow of magnetic irregularities (corresponding to a
constant galactic wind) in the $z$ direction, we find
\begin{equation}
      \label{chic 0}
       \frac{\partial f}{\partial t}
       = -V_c\nabla f+
        \nabla (K\nabla f)+
       \frac{1}{p^2}\frac{\partial }{\partial p}
       \left[K_{pp}\: p^2 \frac{\partial f}{\partial p} \right]
       +Q(r, p, t)
     \end{equation}
which is the  propagation equation we used
in Appendix~\ref{sec: sol diffusion model} (up to several nuclear terms,
cf eq.~(\ref{EQUATION GENERALE})).
In this theory, the momentum diffusion coefficient $K_{pp}$ is naturally
connected to the space diffusion coefficient $K$ by the relation
\begin{equation}
K_{pp}=\frac{{V_a}^2}{9K}p^2
\end{equation}
where $V_a$ is the Alfv\'en speed of the scattering centers.

We emphasize that expressions~(\ref{detailed balance approx})
and~(\ref{chic 0}) are just simplifications of the full Fokker--Planck
equations. As pointed out by Ostrowski \& Siemieniec-Ozi\c{e}blo (1997), if this
simplification is natural in the kinetic description of a gas of
scattering particles, this is no longer the case for particles
scattered off external `heavy' scattering centers.
Nevertheless, this approximation which arises in a  number of astrophysical
cases is sufficient for our purpose. We thus have to treat a simple
energy diffusion
coefficient, which we evaluated in the no--recoil
hard sphere scattering centers approximation as
    \begin{equation}
K_{EE}\equiv \frac{1}{2}
	\left< \frac{(\Delta E)^2}{\Delta t}\right>
	=\frac{2}{3}{V_a}^2\frac{\sigma_Kn_K} c E^2\beta^3
\end{equation}
where $\sigma_Kn_K$ describes the rate of collisions with scatterers which
is related to the mean free path by $\lambda_K=1/\sigma_Kn_K$.
As $K(E)=(1/3)\lambda_K v$, the previous relation
gives the final coefficient to be used in our calculation
   \begin{equation}
   \label{All is for the butter}
K_{EE}=\frac{2}{9}{V_a}^2\frac{E^2\beta^4}{K(E)}
\end{equation}

Our approch is basically similar to that of Heinbach \& Simon (1995).
The meaning of the Alfv\'enic speed derived in their analysis
is not very clear, since the numerical factor in front of
expression~(\ref{All is for the butter}) may vary according to the hypotheses
made for the scattering process.
The assumption that reacceleration occurs only in the thin disc
  is supported by recent complete
magnetohydrodynamics simulations (Ptuskin et al. 1997) describing the evolution
of the hot gas, cosmic rays and magnetic field.

           \subsubsection{Integration into equation}
We define
\begin{equation}
\label{LOSSES}
b^j_{loss}(E)=\left< \frac{dE}{dt}\right>_{Ion}+
       \left< \frac{dE}{dt}\right>_{Coul} 
       +\left< \frac{dE}{dt}\right>_{Adiab}
\end{equation}
and with $K_{EE}^j(E)$ given by (\ref{All is for the butter}),
one obtains (see Appendix)
\begin{equation}
\label{robert}
       A^j_i N^j_{i}(0)=\bar{{\cal Q}^j} -
        2h \frac{\partial}{\partial E}
        \left\{ b^j_{loss}(E)N^j_i(0) -
	K^j_{EE}(E)\frac{\partial}{\partial E}
        N^j_i(0)\right\}
        \end{equation}
This is a second order equation which is solved for each 
order $i$ of the Bessel decomposition.
$A^j_i$ is given by (\ref{Si Ai}) and $\bar{{\cal Q}^j}$ (\ref{SOURCES})
is a generic source term (see Appendix~\ref{sec: sol diffusion model}
for details).
Relation~(\ref{robert}) has been developed on our energy array
so as to be transformed into a matrix equation that is directly inverted.
Two boundary conditions need to be implemented. The effects of energy
losses and diffusive reacceleration have been assumed to be negligible
at the high energy tip ($100\unit{GeV/n}$) of the spectrum.
We have furthermore imposed a vanishing curvature
($\partial^2N_i^j/\partial E^2=0$) at the lowest energy point.
\subsubsection{Galactic wind}
          \label{sec: convection}
It has been recognized for a long time that a
thin disc configuration
would be disrupted by cosmic ray pressure
(Parker 1965, 1966; Ko, Dougherty \& McKenzie 1991).
It can be stabilized by the presence of a halo, but 
further considerations imply that this halo would not be
static either (presence of a convective wind).
Consequences of a wind has been firstly investigated by
Ipavitch (1975) and since then it has been observed in other
galaxies (Lerche \& Schlickeiser 1982a, 1982b; 
Reich \& Reich 1988; Werner 1988). In our own, its effects on nuclei have been
investigated in various models (Jones 1979; K\'ota \& Owens 1980;
Freedman \& al. 1980; Bloemen \& al. 1993),
and in this work we adopted a very simple and tractable
form for $V_c$ throughout the diffusive volume
($dV_c/dz=0$), following Webber, Lee \& Gupta (1992).
A constant wind with free escape boundary is a reasonable approximation to
a more sophisticated variable--wind model (see also conclusions 
from magnetohydrodynamics simulations (Ptuskin et al. 1997)).

		\subsubsection{Comparison with other works}
Our semi--analytical two--dimensional model lies, in some sense, between
the complete numerical resolution of Strong \& Moskalenko (1998) 
and the recent study of Jones et al. (2001).
As compared to the first one, we have a simplified
description of the matter distribution, for which the full analysis of
the parameter space does not require too much {\sc cpu} time.
The second approach can be considered as the limiting case for
our two--dimension model
when the radial extension is much larger than the halo size ($R\gg L$).
Note that a similar analysis was performed by
 Letaw, Silberberg \& Tsao (1993)
in the framework of a leaky box model or Seo \& Ptuskin (1994)
in a one--dimensional diffusion model
but their less systematic approach
does not allow  a full exploration of parameter space.

\section{Data Analysis}
For the aim of our analysis, we can consider different classes
of flux ratios:  primary--to--primary ({\em e.g.} C/O),
secondary--to--primary ({\em e.g.} B/C or sub--Fe/Fe), 
secondary--to--secondary ({\em e.g.} Li/B or Be/B),
ratio of isotopes either stable ({\em e.g.} $^{10}{\rm B}/^{11}{\rm B}$) or
unstable ({\em e.g.} $^{10}{\rm Be}/^{9}{\rm Be}$).
Each of these may be an indicator of some dominant
physical phenomenon and be particularly sensitive to the corresponding
diffusion parameters. The ratio of two primaries
is  practically insensitive to changes in all the parameters,
since they have the same origin and undergo the same
physical processes. So, it can be a very useful tool to fix their
source abundances. We will return to this subject.
Similar conclusions, even if less strong,  may be drawn
for the ratio of two isotopes of the same species, such as
$^{10}{\rm B}/^{11}{\rm B}$.
Indeed, at very low energy values this quantity is slightly affected
by changes in the injection spectra, but the effect is too weak
to constrain free parameters.

The most sensitive quantity is B/C,
as B is purely secondary and its main progenitor C is primary.
The shape of this ratio is seriously modified by changes in the diffusion
coefficient, in the height of the diffusion halo, in the convective
and alfv\'enic velocities.
Moreover, it is also the quantity
measured with the best accuracy, so that it is ideal to test models.
Indeed, being the ratio of two nuclei having similar Z, it is less 
sensitive to systematic
errors than single fluxes or other ratios of nuclei with more distant Z.
For the same reasons, the sub--Fe/Fe may also be useful.
Unfortunately, since existing data are still affected by sizeable
experimental errors, we can only use them to cross--check the
validity of B/C but not to further constrain the
parameters under scrutiny.
Another particularly interesting
quantity is the ratio $^{10}{\rm Be}/^{9}{\rm Be}$. Since
$^{10}{\rm Be}$ is a radioactive element, it is very sensitive to the
processes which can occur in the halo. Therefore, we could use this
ratio in particular to bracket the size of the halo (F. Donato \&
al., in preparation). 


        \subsection{Solar modulation}
We have estimated the effect of the solar wind on the energies and intensities
of cosmic--rays following the prescriptions of the force field approximation
(Perko 1987). The modification in the total interstellar energies
of a nucleus with charge $Z$ and atomic number $A$,
corresponds to the shift
$Z\phi$ (we set the electronic charge equal to unity):
\begin{equation}
E^{\rm TOA} / A \; = \;
E^{\rm IS} / A \, - \, \left| Z \right|  \phi / A.
\end{equation}
Here $E^{\rm TOA}$ and $E^{\rm IS}$ correspond to the top--of--atmosphere
(modulated) and interstellar total energy, respectively.
The solar modulation parameter $\phi$ has the  dimension of a rigidity
(or an electric potential), and its value varies according the
11--years solar cycle, being greater for a period of maximal solar activity.
Often people refer to the equivalent quantity
$\Phi=\left|Z\right|\phi/A\simeq \frac{1}{2}\phi$.
     Once the momenta at the Earth $p^{\rm TOA}$ and at the boundaries
of the heliosphere $p^{\rm IS}$ are determined, the interstellar flux of the
considered nucleus is related to the {\sc toa} flux according to
 the simple rule:
\begin{equation}
\frac{\Phi^{\rm TOA} \left( E^{\rm TOA} \right)}
{\Phi^{\rm IS}  \left( E^{\rm IS} \right)}
\; = \;
\left\{ \frac{p^{\rm TOA}}{p^{\rm IS}} \right\}^{2}
\end{equation}
The determination of the modulation parameter $\Phi$ is totally
phenomenological and suffers from some uncertainty.
As explained in the following, we will deal with data taken around period of
minimal solar activity, for which we fixed $\Phi=250\unit{MV}$.


        \subsection{The dataset we used}
\label{sec:dataset}

In order to test our diffusion model we have widely employed the data 
taken by the
experiment which was launched on board the {\sc nasa} {\sc heao}--3
satellite (Engelmann et al. 1990).
The relative abundances of elements, with charge from 4
to 28 and for energy ranging from 0.6 to 35 GeV/n, have been measured with
unprecedent accuracy.
Data have been  taken in 1979--80,  around a  solar minimal activity.
For the case of B/C, {\sc heao}--3 quoted errors are 2--3$\%$.
To perform the analysis, we also included  data from balloons
(Dwyer \& Meyer 1987) and from the {\sc isee}--3 experiment 
(Krombel \& Wiedenbeck 1988), even if
the relevant error bars are wider.
The first one  collected data in 1973--75 for energies spanning from
around 1.7 to 7  GeV/n.
The second experiment was operating during 1979--81 on board a spacecraft,
in the energy range 100--200 MeV/n.
In some of the figures  presented below, we also
plot -- for purely illustrative goals -- the data point from
{\sc imp}--8 (Garcia--Munoz et al. 1987)
 and the {\sc voyager} experiments (Lukasiak, McDonald \&
Webber 1999)
(we did not include Ulysse data point (DuVernois \& al. 1996), 
since it corresponds to a period
of maximal solar activity). They were not included in the $\chi^2$ analysis
since they are the most sensitive to solar modulation. Nevertheless, we have
checked that the best $\chi^2$ values were not significantly
modified when these points were added.

As regards the sub--Fe(Sc+V+Ti)/Fe ratio, we used
data from {\sc heao}--3 (Engelmann et al. 1990) and from balloons
(Dwyer \& Meyer 1987).
In both cases the error bars,  around 10$\%$, are
significantly larger than for B/C.


        \section{Results}
        \label{sec: results}   
     We varied the relevant parameters $K_0$, $L$, $V_c$, $V_a$, 
     and $\delta$ of the diffusion model described above.
       The constraints are much simpler to express with the combinations
       $L$, $K_0/L$, $V_c$ and $V_a/\sqrt{K_0}$  (this last expression
       appears naturally in $K_{EE}(E)$ describing reacceleration).
        As for the primaries, we have adjusted the source abundance for
       nitrogen which is not a pure secondary.
      We  calculated the $\chi^2$ over the
      26 experimental points listed in \ref{sec:dataset}, for each possible
      combination obtained varying the free parameters in the whole 
      parameter space.
     In the following, we will focus on B/C and we cross--check 
     the compatibility of the
     obtained parameters when applied to other fluxes.     
      
         \subsection{Results of B/C  and sub--Fe/Fe analysis for $\delta=0.6$}
In a first step, the power $\delta$  of rigidity in the expression
      for the diffusion
      coefficient (see Eq. (\ref{eq:diffusion_coefficient}))
      has been fixed to 0.6.
      This will be the reference value for the bulk of our analysis.
      We only show models giving a $\chi^2$ less than 40.
      The best models have $\chi^2 \approx 28$. Some of these configurations
      have a very small halo size in which case the
      condition $h\ll L$ is no longer valid. Thus we also required
      $L>1\unit{kpc}$ in the whole analysis.

\begin{figure}[p]
\plotone{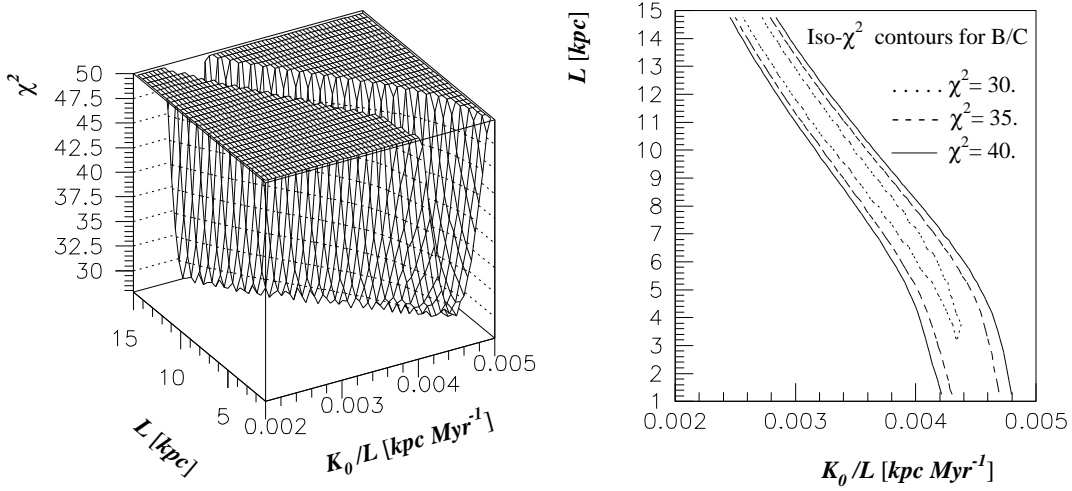}
\caption{The $\chi^2$ (adjustment to B/C data, see text) has been
computed in all the parameter space for $\delta=0.6$ (defined by
$K=K_0{\cal R}^\delta$).
A best $\chi^2$ is obtained for each $L$ and $K_0/L$. 
Left figure displays the $\chi^2$ values in the $K_0/L$ -- $L$ plane. It is
truncated for $\chi^2>50$. Right figure shows the
iso--$\chi^2$ lines of the previous surface.}
\label{fig2}
\end{figure}

\begin{figure}[p]
\plotone{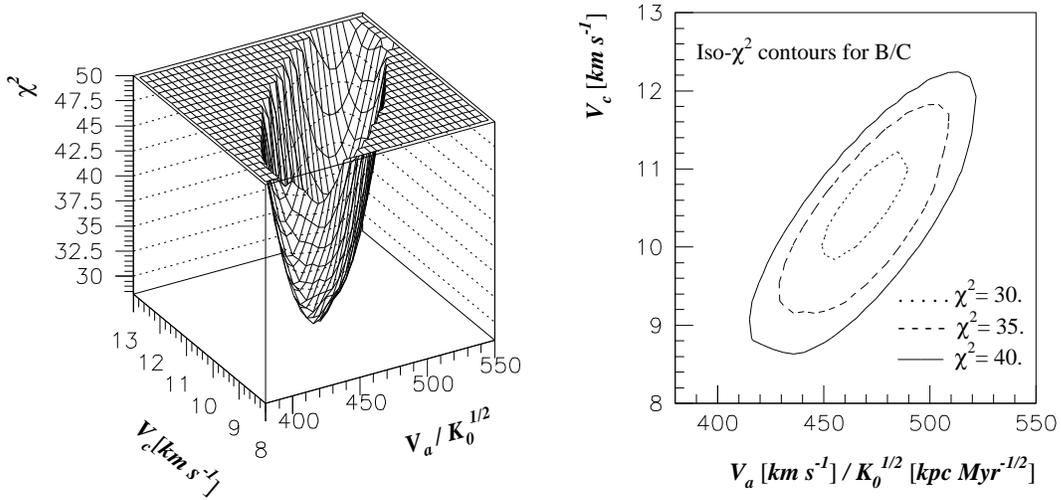}
\caption{As in Fig. \ref{fig2},
a best $\chi^2$ is obtained for each $V_c$ and $V_a/\sqrt{K_0}$.
Left and right figures are similar to those in Fig.~\ref{fig2}.}
\label{fig3}
\end{figure}

      The first result  is that only the source spectrum
      $q^j(E) = q_0^j \, p^{-\alpha_j}$ gives a good fit to the data.
      It will be used all the way long.
      For each $L$ and $K_0/L$, we vary $V_a/\sqrt{K_0}$ and $V_c$ and the best
      $\chi^2$ are shown in Fig.~\ref{fig2}. All the good models are
      located in a narrow strip in the plane $L$ -- $K_0/L$, showing a strong
      correlation between these two parameters.
      However, there is still a degeneracy.
      It should be lifted with further specific
      analysis of radioactive nuclei (F. Donato \& al., in preparation).
      Conversely, for each $V_a/\sqrt{K_0}$ and $V_c$ we varied
      $L$ and $K_0/L$ and the best  values for the $\chi^2$ are depicted
      in Fig.~\ref{fig3}. We see that they are gathered
      in a narrow range, namely: $8.5\lesssim V_c \lesssim 12 $ km sec$^{-1}$ and
      $410 \lesssim V_a/\sqrt{K_0} \lesssim 530 \unit{km}\unit{s}^{-1}
      /\unit{Mpc}\unit{Myr}^{-1/2}$.
      We do not find any model having a good $\chi^2$
      without convection ($V_c=0$) or without reacceleration ($V_a=0$).
      We emphasize the fact that these four parameters are strongly correlated
      so that, when one of them is determined (for example $L$
      with radioactive
      nuclei), the allowed ranges for the others will be narrower than
      what could naively appear from the above two figures.

\begin{figure}[p]
\plotone{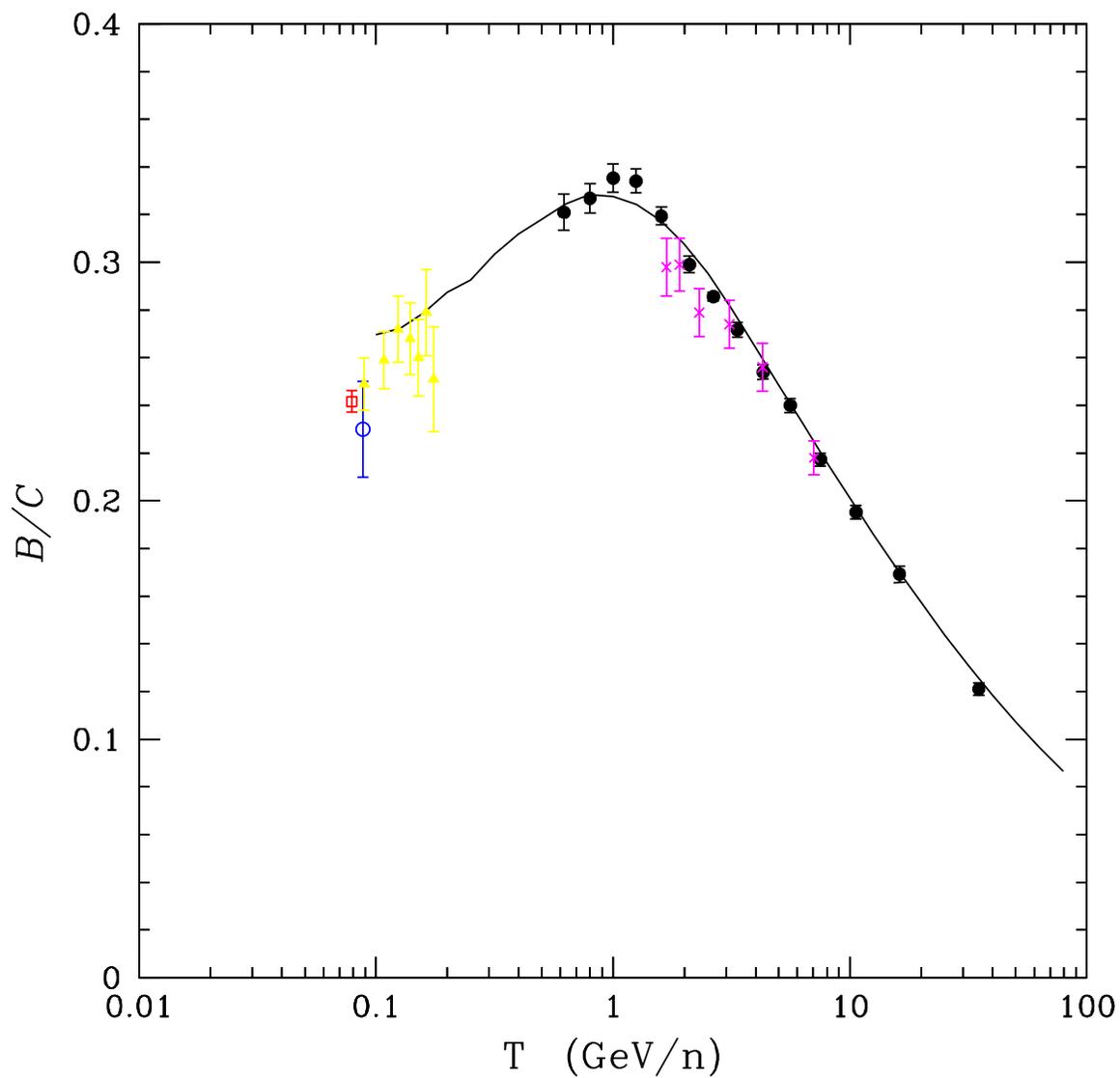}
\caption{This curve displays the computed ratio of 
($^{10}$B+$^{11}$B)/($^{12}$C+$^{13}$C+$^{14}$C) for a
configuration giving a reduced $\chi^2_r \approx 1.3$. The experimental points
are from {\sc heao}-3 (solid circles), {\sc isee} (triangles), 
{\sc imp}-8 (empty circle), {\sc voyager} (square) and balloons (crosses).}
\label{fig1}
\end{figure}

\begin{figure}[p]
\plotone{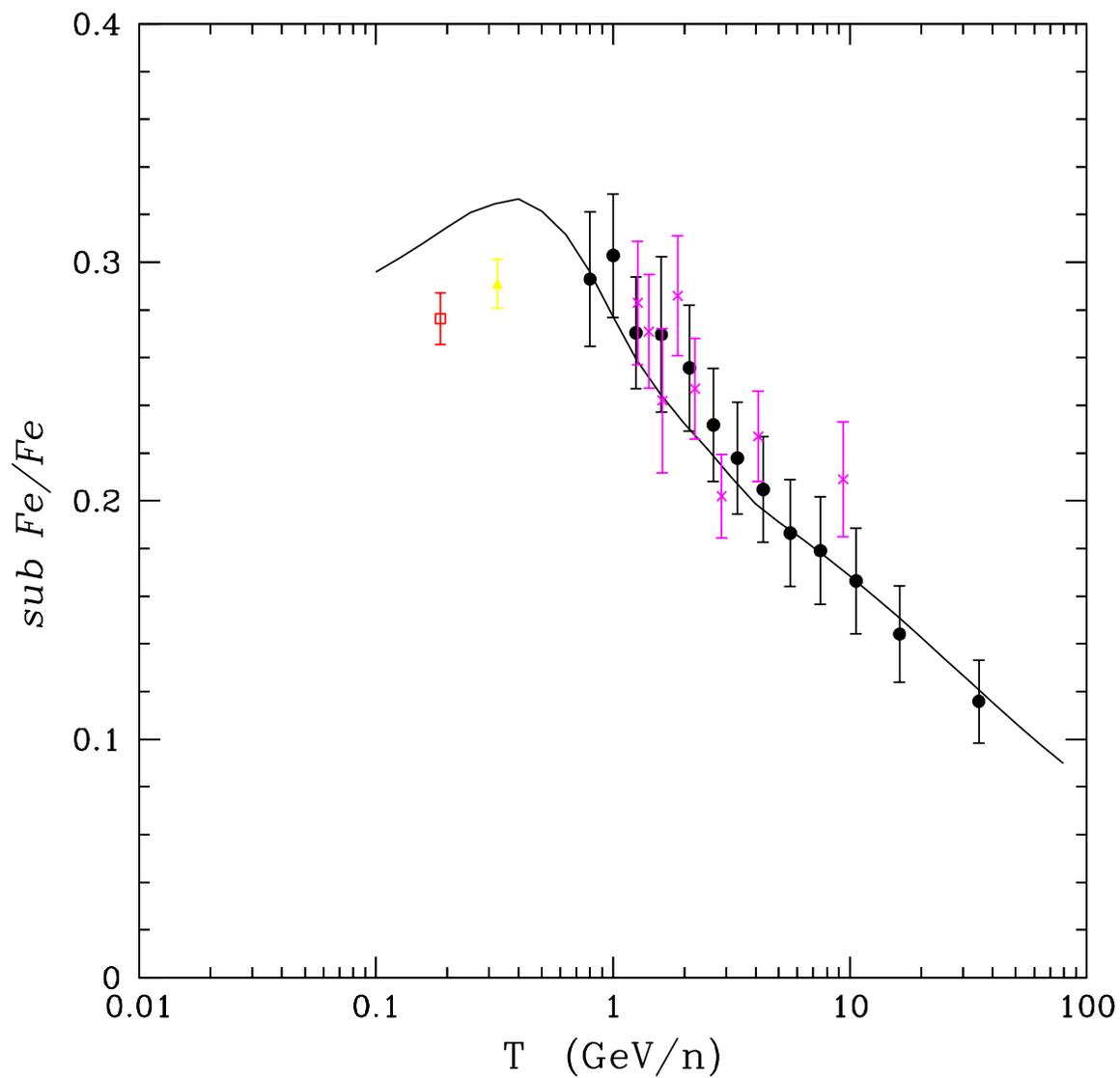}
\caption{This curve displays the computed ratio of (Sc+Ti+V)/Fe for
the same configuration as in Fig. \ref{fig1}. The experimental points
are from {\sc heao}-3 (solid circles), {\sc isee} (triangles),
 {\sc voyager} (square) and balloons (crosses).}
\label{fig1bis}
\end{figure}

      The observed sub--Fe/Fe flux ratio does not give very powerful 
constraints.
      Nevertheless, all the models which give a good fit to B/C data
are consistent with sub--Fe/Fe data.
      Figures~\ref{fig1} and \ref{fig1bis} illustrate
      these results with the particular choice $L=9.5 \unit{kpc}$,
      $K_0/L=0.00345 \unit{kpc}\unit{Myr}^{-1}$, $V_c=10.5\unit{km}\unit{s}^{-1}$
      and $V_a/\sqrt{K_0}=470\unit{km}\unit{s}^{-1}
      /\unit{Mpc}\unit{Myr}^{-1/2}$, giving a reduced $\chi_r^2\approx 1.3$.
      Note that all the configurations lying in the allowed ranges of Fig.
      \ref{fig2} and \ref{fig3} will provide very similar curves.

\begin{figure}[p]
\plotone{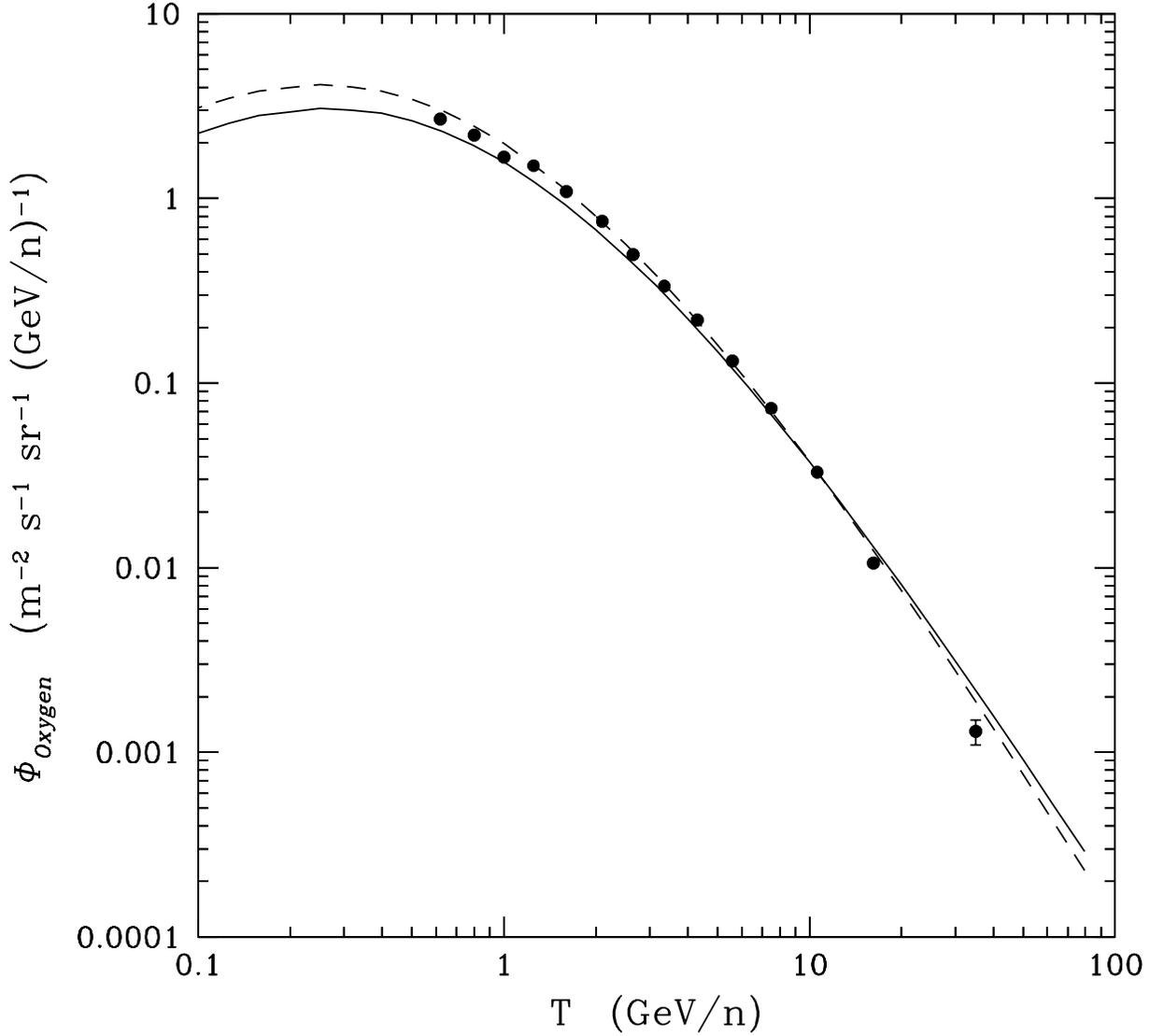}
\caption{Theses curves display the computed flux of Oxygen for
the same configuration as in figure \ref{fig1}. The solid line corresponds 
to $\delta + \alpha_{Oxygen}=2.68$, the dotted line 
to $\delta + \alpha_{Oxygen}=2.80$ (see text for details). 
The experimental points are from {\sc heao}-3.}
\label{fig1ter}
\end{figure}

 Up to now we have dealt with flux ratios, which partially hide the spectral
features of the fluxes themselves. So we have also checked
that the above configurations were consistent
with the oxygen flux as measured by {\sc heao}--3.
In Fig.~\ref{fig1ter} the oxygen flux has been calculated for 
the same configuration as in Fig.~\ref{fig1} and \ref{fig1bis}. We can 
notice that the 
agreement with  {\sc heao}--3 data is not so satisfactory. The parameter 
which mainly prevents us from obtaining a good fit is the source spectrum. 
We recall that we set the sum $\delta + \alpha_j$ 
to the power--low index as determined in Wiebel--Sooth et al.
(1998). 
Indeed, this number for the oxygen species is equal to $2.68$ as derived
from {\sc heao}--3 and many other measurements.
If instead we fix $\delta + \alpha_{Oxygen}$ to be $2.8$
(as in Swordy \& al. (1993)), the fit improves (see Fig. \ref{fig1ter}).

         \subsection{Variation of the diffusion coefficient spectral index}
\begin{figure}[p]
\plotone{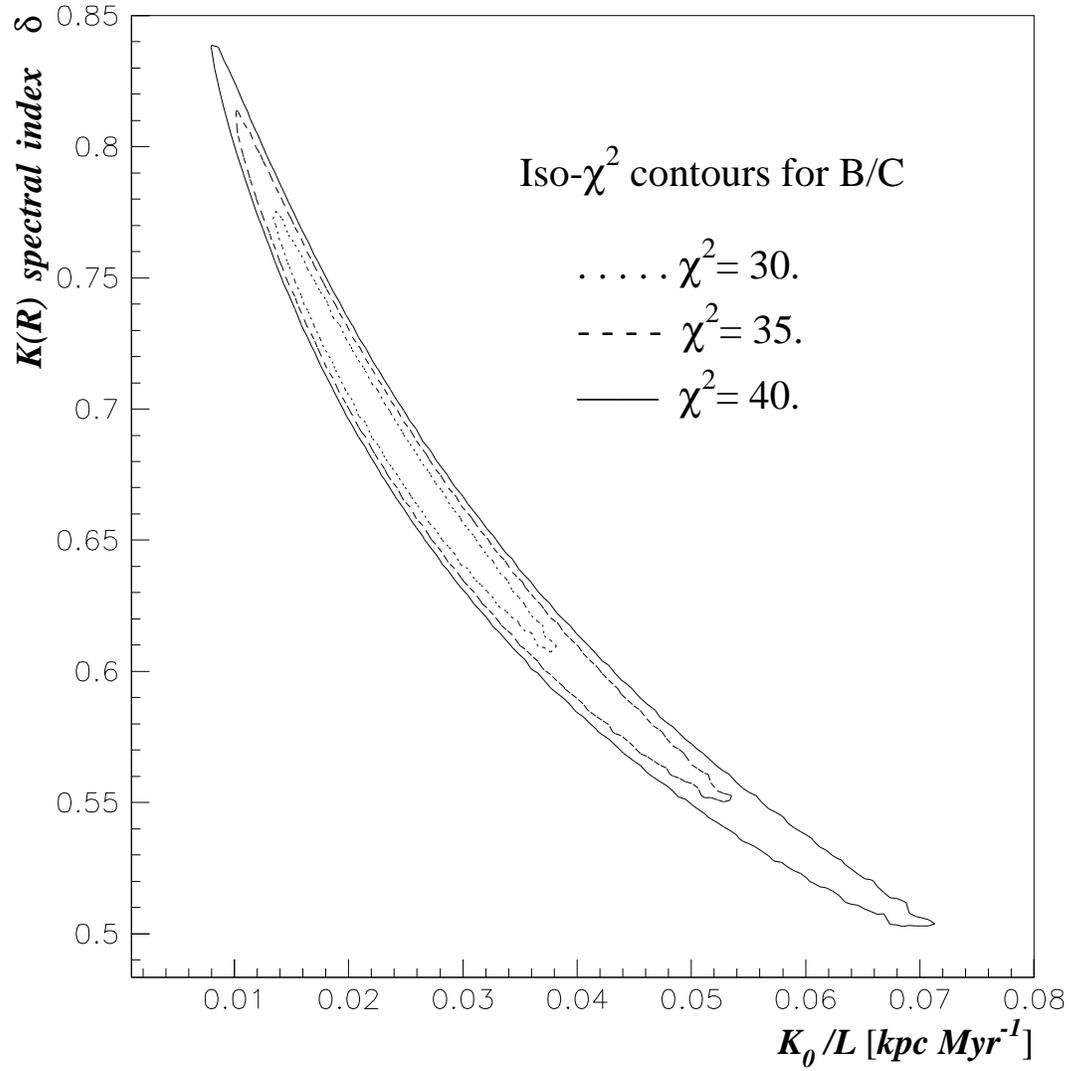}
\caption{The parameter space has been explored  
for a fixed value of the halo thickness $L=3 \unit{kpc}$, 
varying $\delta$
(spectral index of $K$) as well as the other parameters.
A best $\chi^2$ is obtained for each $\delta$ and $K_0/L$.
This figure shows the
contour lines for fixed values of $\chi^2$.}
\label{sctroumph}
\end{figure}

\begin{figure}[p]
\plotone{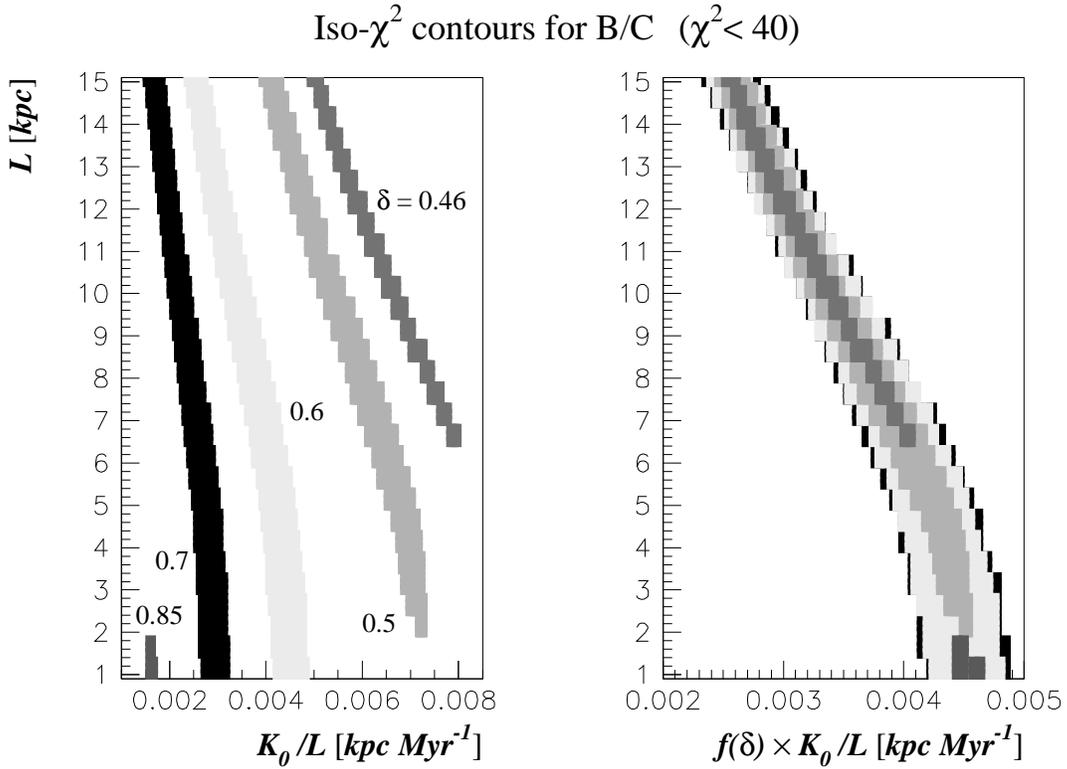}
\caption{Models with different values of $\delta$ are shown. 
As in the previous figures, for each value of $L$ and $K_0/L$,
only the best $\chi^2$ value is retained when the other parameters $V_c$ and
$V_a/\sqrt{K_0}$ are varied.  
The figure in the left panel displays the contour levels for
$\chi^2 < 40$ for the indicated values of $\delta$. It is possible to scale the 
$K_0/L$ values by a function $f(\delta)$ to superimpose the contours corresponding
to different values of $\delta$ (see text). This is displayed in 
the right panel.}
\label{sctroumphette}
\end{figure}

We also tested different values for $\delta$, and we find that 
correspondingly the diffusion
parameters giving a good fit to B/C change.
As an example, for a fixed value of the halo 
thickness $L=3 \unit{kpc}$,
we find that $\delta$ is allowed to vary between $0.5$ and $0.84$.
This is displayed in Fig.~\ref{sctroumph}.
In the whole parameter space, the range of $\delta$ extends
from approximately $0.45$ to $0.85$, as one can see in Fig. 
\ref{sctroumphette}. In particular the value $\delta = 0.33$ 
corresponding to a Kolmogorov--like turbulence spectrum is strongly 
disfavoured ($\chi^2 > 100$).
For intermediate values of $\delta$, good models are obtained for 
the full range in $L$ (as in Fig. \ref{fig2}).
For low values of $\delta$, models with a small halo size $L$ are excluded;
in particular
for $\delta < 0.45$, there is no good model with $L<15 \unit{kpc}$.
Finally, for high values of $\delta$, models with a large halo $L$ are excluded, and
for $\delta > 0.85$, there is no good model with $L>1 \unit{kpc}$. 
As can be seen in Fig. \ref{sctroumphette}, each value of $\delta$ 
gives a different contour plot in the $K_0/L$ -- $L$ plane. It appears that 
they all can be superimposed to a single curve by a rescaling $K_0/L \rightarrow K_0/L
\times f(\delta)$, where $f$ is a function of $\delta$ only. 
For the contours displayed in Fig. \ref{sctroumphette}, it takes the values
$f(0.46)=0.51$, $f(0.5)=0.62$, $f(0.6)\equiv1$, $f(0.7)= 1.54$ and 
$f(0.85)=2.78$. 

In the $V_a/\sqrt{K_0}$ -- $V_c$ plane, the values of $V_c$ are shifted 
downward as $\delta$ is decreased but the allowed range of $V_a/\sqrt{K_0}$
does not significantly move. Nevertheless, the allowed values for
$V_c$ never reach $0$, so that no--wind models can be excluded.

\section{Discussion and conclusion}
We obtain good quantitative constraints on the diffusion parameters from
B/C data. 
In particular, there is a very strong correlation between $L$,
$K_0/L$, $V_c$, $V_a/\sqrt{K_0}$ and $\delta$. For $\delta=0.6$,
we find that $8.5 \unit{km}\unit{s}^{-1}<V_c<12 \unit{km}\unit{s}^{-1}$
and $410 < V_a/\sqrt{K_0} < 530$ (where $V_a$ is expressed in
$\unit{km}\unit{s}^{-1}$ and $K$ in $\unit{kpc}^2\unit{Myr}^{-1}$).
Furthermore, we show that the power law index for the diffusion coefficient is
restricted to the interval $[0.45, 0.85]$, the best $\chi^2$
being $25.5$ for $\delta=0.70$.
For any $\delta$ in this interval, the good parameters in the
$K_0/L$ -- $L$ and $V_a/\sqrt{K_0}$ -- $V_c$ planes can be straightforwardly
deduced from the corresponding values for $\delta=0.6$ by a simple scaling 
law. We exclude any model without a convective velocity or without
reacceleration for any combination of the three other diffusion parameters.

Our conclusions could get more stringent by new measurements in
the whole energy spectrum for all nuclei.
We emphasize that all our results were obtained using the best data, which
are rather scarce and more than 20 year--old;
new data are thus strongly needed. The {\sc ams} experiment
on board the International Space Station 
will have in principle the ability to provide some of these data.

The next steps of this analysis will be to study radioactive species.
In particular,  we expect the recent {\sc smili} data
(giving the spectral distribution of Be isotopes over a large energy range)
to provide a new insight on cosmic ray propagation,
and thus to constraint further the diffusion parameters.
We are also investigating the standard antiproton signal, using the results
described in this paper.
\section*{Acknowledgments}
We thank Aim\'e Soutoul for having pointed out a mistake in an early version
of this paper. D.M. is particularly grateful for his remarks. 
F.D. gratefully acknowledges a fellowship by the Istituto Nazionale di Fisica
Nucleare. We also would like to thank the French Programme National de
Cosmologie for its financial support. 
\newpage
\addtocontents{toc}{\protect\newpage}
\addcontentsline{toc}{part}{Annexes}
\addtocontents{toc}{\protect\vspace{2ex}}

\appendix
\section{Solution for the diffusion model}
\label{sec: sol diffusion model}
Sources and interactions with matter are confined to the thin disk and
diffusion which occurs throughout disc and halo with the same strength
is independent of space coordinates.
Sources follow a universal
form $Q(E)q(r)$ with an associated normalization $q^j_0$ for each nucleus
(see $\S$ \ref{grumph}).
Furthermore, we consider here only constant wind $V_c$ in the $z$ direction
so that adiabatic losses (third term in 
(\ref{EQUATION LA PLUS GENERALE})) vanish.

\subsection{High energy limit}
The dependence in energy $E$ is implicit, and if we add radioactive
contributions localized in both disc and halo
\begin{eqnarray}
       \label{EQUATION GENERALE}
      \left( {\cal L}_{dif\!f} -\Gamma^j_{rad}\right) N^{j}(r,z)+
         \sum_{k=1}^{j-1}\Gamma_{rad}^{kj}N^{k}
         +2h\delta(z) \left(
         q^j_0Q(E)q(r)+\sum_{k=1}^{j-1}\tilde{\Gamma}^{kj}N^{k}(r,0)
         -\tilde{\Gamma}^j N^{j}(r,0)\right)=0\\
       \label{eq: operateur diffusion}
           {\cal L}_{dif\!f} =-V_{c} \frac{\partial}{\partial z}
          +K\left(\frac{\partial^{2}}{\partial z^{2}}+
          \frac{1}{r}\frac{\partial}{\partial r}
(r\frac{\partial}{\partial r})\right)
          \hspace{4.5cm}
       \end{eqnarray}
In our geometry a solution can be found in term of Bessel functions of
zeroth order taking advantage of their properties (Jackson 1975). One can expand
all the quantities over the orthogonal set of Bessel functions 
$\{J_0(\zeta_i x)\}^{i=1\dots \infty}$,
\begin{eqnarray}
N^{j}(r,z)=\sum_{i=1}^{\infty}N^j_{i}(z)J_{0}(\zeta_{i}\frac{r}{R})\hspace{3cm}\\
      q(r)=\sum_{i=1}^{\infty}\hat{q_i}J_{0}(\zeta_{i}\frac{r}{R}),~~~~~~~~
      \hat{q_{i}}=\frac{1}{\pi R^{2}J^{2}_{1}(\zeta_{i})}
                    \frac{\int_{0}^{1}\rho q(\rho)J_{0}(\zeta_{i}\rho)d\rho}
                     {\int_{0}^{1}\rho q(\rho)d\rho},~~~~~~~\rho\equiv r/R
\end{eqnarray}
Such an expression respects automatically one of the boundary condition
$N(r=R,z)=0$.
We will use a compact notation to describe the most general form for a
source term
\begin{equation}
\label{SOURCES}
\bar{{\cal Q}^j}=q_0^jQ(E)\hat{q_i}+\sum_k^{m_k>m_j}
            \tilde{\Gamma}^{kj}N_i^{k}(0)
\end{equation}

\paragraph{Stable progenitors \boldmath{$N^k$}} As equations have to be
valid order by order (second term of~(\ref{EQUATION GENERALE}) vanishes)
\begin{equation}
        \label{PRIM i}
        \left[ \frac{d^{2}}{dz^{2}}-\frac{V_{c}}{K}
            \frac{d}{dz}- (\frac{\zeta_{i}^{2}}{R^{2}}+
            \frac{\Gamma_{rad}^{N^{j}}}{K})\right] N^{j}_{i}(z) =
            \left( -\frac{\bar{{\cal Q}^j}}{K} +\frac{2h\tilde{\Gamma}^j}{K}
	    N^{j}_{i}(0)\right)\delta (z)
         \end{equation}
A fast procedure to solve~(\ref{PRIM i}) is given by the following three steps:
       \begin{enumerate}
         \item Solution in the halo (no right--hand side in eq.~(\ref{PRIM i}))
         with the boundary condition $N^j_i(z=\pm L)=0$
         \item Integration $(\lim_{h\rightarrow 0}\int_{-h}^{+h}\dots dz)$ of
         equation~(\ref{PRIM i}) through the thin disc\footnote{
         In terms of distribution (quoted in braces), defining $\sigma_0$
and $\sigma_1$ as the
         discontinuities of $0^{th}$ et $1^{st}$ order, remember that
           \[ \textstyle
          \frac{\partial^2}{\partial z^2} \{ {\cal F}(z) \} = \left\{
              \frac{\partial^2{\cal F}(z)}{\partial z^2}\right\}
+\sigma_{1}\delta (z)+
               \sigma_{0}\frac{\partial \delta (z)}{\partial z}
           \]
      Imposing the continuity of the vertical cosmic ray current across the plane
      $z=0$, we thus have\\
            $\textstyle \sigma_1 \equiv lim_{\epsilon\rightarrow 0} \left[
              d N^j_i (z)/dz\right]^{+\epsilon}_{-\epsilon}=-2
              N^j_i(0)\frac{V_{c}}{K}$ and $\sigma_0=0$.
            }, which gives
              \begin{equation}
             \label{DISQUE}
             \textstyle 2 {N'}^j_i (z)|_{z=0}-2 N^j_i (0)\frac{V_c}{K}-2h N^j_i
             (0)\frac{\tilde{\Gamma}^j}{K}+\bar{{\cal Q}^j}=0
             \end{equation}
         \item Put the halo solution in equation~(\ref{DISQUE})
         to ensure continuity beetwen the two zones.
       \end{enumerate}
We finally obtain the solutions for stable progenitors in relativistic regime:
\begin{eqnarray}
        \label{SOL PRIMAIRES}
      N^j(r,z)=\exp\left( \frac{V_{c}z}{2K}\right)~
            \sum_{i=0}^{\infty}~ {\frac{\bar{{\cal Q}^j}}{A^j_i}}~
            \frac{\sinh\left[ \frac{S^j_i (L-z)}{2} \right]}
                 {\sinh\left[ \frac{S^j_i L}{2} \right]}~ J_0 
		 (\zeta_i\frac{r}{R})
            \end{eqnarray}
\begin{eqnarray}
       \textstyle \bar{{\cal Q}^j}\equiv q_0^jQ(E)\hat{q_i}+\sum_k^{m_k>m_j}
            \tilde{\Gamma}^{kj}N_i^{k}(0)\hspace{3.cm}\\
       \textstyle \label{Si Ai}
         S^j_{i}\equiv ( \frac{V_{c}^{2}}{K^{2}}
         +4\frac{\zeta_{i}^{2}}{R^{2}} +4\frac{\Gamma^{N^{j}}_{rad}}{K} )
         ^{1/2} \hspace{5mm}
        ~~~~~~~~~~A^j_i \equiv 2h\tilde{\Gamma}^{tot}_{N^{j}}+V_{c}+K S^j_{i}
            \coth ( \frac{S^j_{i}L}{2})
      \end{eqnarray}
For a primary $\bar{{\cal Q}^j}= q_0^jQ(E)\hat{q_i}$, and for a pure secondary
$\bar{{\cal Q}^j}=\sum_k^{m_k>m_j} \tilde{\Gamma}^{kj}N_i^{k}(0)$.
Note that solutions given  in Webber et al. (1992)
for secondary takes
advantage of the primary form of $N_i^{k}(0)$. Since we are here interested
in a {\em shower--like} (see $\S$ \ref{sec: spallation reaction}) 
resolution, the form given here is more adapted.

\paragraph{\boldmath{$\beta$} decay contribution from \boldmath{$N^k$}}
For all the nuclei treated here, $N^j$ never has more than one unstable
contribution, so that the sum over $k$ for $N^k_{rad}$ reduces to one term
in equation~(\ref{EQUATION GENERALE}). Resolution is complicated by the
localisation of this source in the whole halo. Focalising on this
specific term,
neglecting for a while primary source and {\em classical} spallative secondary
contribution $2h\delta(z) \sum_k\tilde{\Gamma}^{kj}N^{k}(r,0)$, one obtains
(following the same procedure as described in
the previous section\footnote{The contribution
of these radioactive nuclei may be unimportant in some cases, but we should
take it into account as it is the dominant process for some others. In the
simple example of $^{10}$Be$\rightarrow^{10}$B, neglecting this channel
would give
an error of about $10\%$ on the B flux, whereas considering 
that this term is only located in the disc would give an error 
of about $3\%$ compared to the rigourous
treatment given above. Notice finally that at fixed energy per nucleon, the
rigidity depends on the nuclear species at stake.
The diffusion coefficient $K^j$ of the child nucleus $j$ is therefore
different from its progenitor's one $K^k$.
The difference $K^j-K^k$ tends to vanish for the heaviest nuclei.})
\begin{eqnarray}
       \label{SOL SECONDAIRES RAD}
        N^j_{\Gamma^k_{rad}} (r,z)=& \displaystyle\sum_{i=0}^{\infty}J_0
(\zeta_i \frac{r}{R})\times
        \frac{\Gamma_{rad}^{kj}}{{K^j}(a_i^2-a^2)}\:
            \frac{N^{k}_{i}(0)}{\sinh\left( \frac{S^k_i L}{2} \right)}
            \hspace{7.cm} \\
            \times& \left\{
            \begin{array}{ll}
            -\alpha \: a_i \cosh[ \frac{S^j_i (L-z)}{2} ]
            e^{(\frac{V_c z}{2K^j})}+\left( a \sinh[ \frac{S^k_i (L-z)}{2} ]
            +a_i \cosh[\frac{S^k_i (L-z)}{2}] \right) e^{(\frac{V_c z}{2K^k})}
             \vspace{2.mm} \\
            +\nonumber
            \frac{e^{(\frac{V_c z}{2K^j})}}{A^j_i}\frac{\sinh\left[ \frac{S^j_i
             (L-z)}{2} \right]}{\sinh\left[ \frac{S^j_i L}{2} \right]}
              \left[
              \begin{array}{lll}
             \textstyle  \vspace{2mm} \alpha\:a_i \cosh\left[\frac{S^j_i
L}{2}\right]
             \left( V_c +
             2h\tilde{\Gamma}^j+K^j S^j_i \tanh\left[\frac{S^j_i
             L}{2}\right]\right) \\
            \textstyle \vspace{2mm}+\sinh\!\left[\frac{S^k_i
L}{2}\right]\left[a\left(\frac{V_c
             K^j}{K^k}-2V_c-2h\tilde{\Gamma}^j\right)-a_i K^j
             S^k_i\right]\\
             \textstyle
             +\cosh\left[\frac{S^k_i L}{2}\right]\left[a_i\left(\frac{V_c
             K^j}{K^k}-2V_c-2h\tilde{\Gamma}^j\right)-a K^j S^k_i\right]
              \end{array}
             \right]
            \end{array}
            \right\}
       \end{eqnarray}

       where $S^j_{i}$ and $A^j_i$ have already been defined in (\ref{Si Ai}) and
       \begin{eqnarray}
         \label{alpha}
         \textstyle \alpha\equiv\exp \left( \frac{V_c
         L}{2}\left(\frac{1}{K^k}-\frac{1}{K^j}\right)\right),~~~
         \label{a}
         \textstyle
         a\equiv\frac{V_c^2}{2K^k}\left(\frac{1}{K^k}-\frac{1}{K^j}
	\right)+\frac{\Gamma_{rad}^{kj}}{K^k}-
	\frac{\Gamma_{rad}^{j}}{K^j},~~~
         \label{a_i}
         \textstyle
         a_i \equiv\frac{S_i^k V_c}{2}\left(\frac{1}{K^k}-\frac{1}{K^j}\right)
         \end{eqnarray}
\subsection{Full solution}
Modifying relation~(\ref{EQUATION GENERALE}) so as to
take into account energy losses and diffusive reacceleration
is straightforward since those processes take place only in the 
disc and not in the halo.
Following step by step the three point
procedure previously described, one gets the differential 
equation\footnote{For the $\beta$--unstable
progenitor, the derivation is a bit more tedious but leads to the same
final compact result.}
\begin{equation}
        \label{SOLUTION NON RELATIVISTE}
         A^j_i N^j_{i}(0)=\bar{{\cal Q}^j} -
        2h \frac{\partial}{\partial E}
        \left\{ b^j_{loss}(E)N^j_i(0) - K^j_{EE}(E)\frac{\partial}{\partial E}
       N^j_i(0)\right\}
        \end{equation}
   where $b^j_{loss}(E)$ and $K^j_{EE}$ respectively correspond 
   to energy losses (described in section~\ref{sec: energy losses}
   and defined by (\ref{LOSSES}))
   and energy diffusion (reacceleration term
   given by~(\ref{All is for the butter})).



\clearpage

\suppressfloats
\newpage

\begin{figure}[p]
\plotone{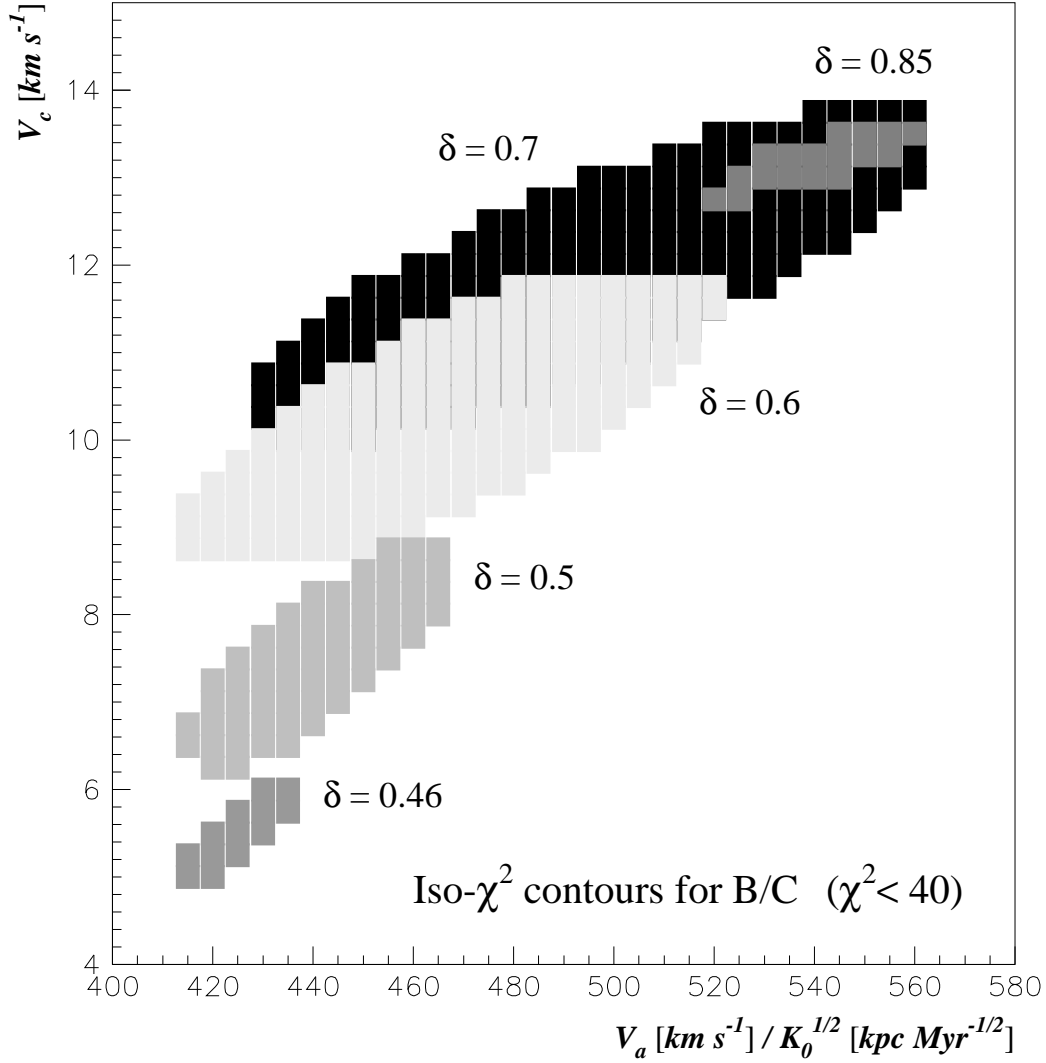}
\caption{Models with different values of $\delta$, 
the coefficient diffusion spectral
index, are shown. For each value of $V_c$ and $V_a/\sqrt{K_0}$ ,
only the best $\chi^2$ value is retained when the other parameters $L$ and
$K_0/L$ are varied.  The figure displays the contour levels for
$\chi^2 < 40$ for the indicated values of $\delta$.}
\label{sctroumphette2}
\end{figure}
\vfill

\end{document}